# Synthetic metal line indices for elliptical galaxies from super metal rich $\alpha$-enhanced stellar models

**Achim Weiss[1], Reynier F. Peletier[2,3] and Francesca Matteucci[2,4]**

[1] Max-Planck-Institut für Astrophysik, Postach 1523, 85740 Garching, Germany
[2] European Southern Observatory, 85478 Garching, Germany
[3] Kapteyn Institute, Univ. of Groningen, The Netherlands
[4] Dep. of Astronomy, Univ. of Trieste, Italy

**Abstract.** There are strong indications from recent papers (e.g. Worthey et al. 1992) that the abundance ratio of Mg/Fe, and consequently also O/Fe in giant elliptical galaxies is not solar. The line strengths of two Fe lines at 5270 and 5335 Å are weaker than one expects from the strength of the Mg b line if [Mg/Fe] = 0. We have synthesized absorption line indices to derive the Mg and Fe abundances of these galaxies. For these models we have calculated new evolutionary tracks of high metallicity stars with a range of Mg/Fe abundances. This is the first time that such tracks have been generated. Integrating along isochrones to synthesize metal line strengths we find that for a typical bright giant elliptical [Mg/Fe] should be between +0.3 and +0.7. We show that this result is independent of other parameters such as age, total metal content and mixing length parameter. The total metal content is super-solar, but the iron metallicity of elliptical galaxies not necessarily should be larger than solar. For the formation of elliptical galaxies our result on the Mg and Fe abundances has the implication that most of the enrichment of the gas has to come from SNe II, which have more massive progenitors and as such produce relatively more O and Mg than Fe. It means that most of the stars have to be formed within a period of $3 \times 10^8$ years, so that there can only be one major collapse phase of the galaxy.

# 1 Introduction

Understanding the metal content of the dominant stellar population in elliptical galaxies is a difficult and long-standing problem. Since we can only study these galaxies in integrated light, and since the velocity broadening forces us to use only strong lines, it has up to now not been possible to uniquely determine the average age and metallicity of elliptical galaxies in general. Most of the information about the composition of elliptical galaxies comes from the line strengths of a few strong absorption lines, of which the $Mg_2$ index, describing the Mg b line at 5180 Å (Faber 1977) is the most important. Other lines that are often used are those of the system defined by Burstein et al. (1984), lines in the near-ultraviolet (Rose 1985) or in the near-infrared (e.g. Carter, Visvanathan & Pickles 1987).

The strength of a line is the sum of the line strengths of the individual stars, weighted with their luminosity in this region of the spectrum. If the overall metallicity of the system increases, higher opacities will result, so that the temperature of the stars will drop, causing stronger strengths for most lines. The same happens if the age of the system is increased. If the abundance of a certain element is enhanced, only the strengths of the lines that are due to that element will increase, when the opacity structure of the stars in general remains unchanged. In short: the strength of an absorption line is determined by many factors, and it is hard to get the metallicity of the system from one line. With additional assumptions however, one can say more about the global metallicity. Assuming that elliptical galaxies are as old as globular clusters, and that the abundance ratios between the various elements heavier than He are the same as in the sun, there is quite a unique relation between $Mg_2$ and metallicity. The first calibration of this relation has been given by Mould (1978) and later others have followed (see e.g. Brodie & Huchra 1991, who derived much lower metallicities than one gets using the calibration by Mould). Assuming a different age, galaxies should obey a different $Mg_2$ – metallicity relation. But whether one chooses an old age, or only $\sim 8$ Gyr (O'Connell 1980, Pickles 1985), one continues to derive metallicities for elliptical galaxies that are solar or larger. The stars in these systems should have a spectral appearance like the strong-lined stars found in the bulge of our Galaxy (Whitford 1978, Rich 1988).

Due to its proximity, people have been able to measure the metallicity distribution of stars in the galactic bulge. Rich (1988) found a metallicity range between [M/H] = –1.0 and 1.0, with a peak at [M/H] = 0.3. Geisler & Friel (1992) have subsequently confirmed Rich's results on a larger sample of bulge stars, with the difference that [M/H] = 0.12 for the peak, which is in closer agreement with the latest value of 0.08 by Rich (1992).

This distribution was theoretically reproduced by Matteucci & Brocato (1990) in a chemical evolution model for our Galaxy. In their model they predicted that the abundance ratios of $\alpha$-elements to Fe (with $\alpha$ we indicate $\alpha$-rich isotopes such as O, Mg, Si etc..) in bulge stars should be larger than solar over almost the full Fe range, in order to reproduce the metallicity distribution found by Rich (1988). Such non-solar ratios are not a novelty. In fact, halo field and globular cluster stars generally have [$\alpha$/Fe] ratios larger than solar (see Wheeler, Sneden & Truran 1989 for a review). It is believed that the $\alpha$-elements in these stars are enhanced because they were formed rapidly in the early phases of Galaxy evolution, when most of the enrichment of the ISM was due to supernovae (SNe) of type II, which produce much more O and other $\alpha$-elements relative to Fe as opposed to SNe of type Ia. These latter, occurring on much longer timescales than SNe II, are thought to affect mostly the evolution of the Galactic disk (Greggio & Renzini 1983; Matteucci & Greggio 1986). The main difference between bulge stars and halo and globular cluster stars is their absolute metal content, being much higher in bulge stars. The reason for bulge stars to show overabundances of the ratio of $\alpha$-nuclei and iron as compared to the sun is, in Matteucci and Brocato's model, the consequence of the fact that the bulge evolved much faster than the solar neighbourhood. This way, during the collapse phase of the galaxy the absolute metallicity increases much faster in the bulge than in the halo and disk, on timescales shorter than the typical timescales for the appearance of type Ia SNe.

Published high dispersion measurements of some bulge stars do not allow any firm conclusion on abundance ratios of bulge stars (Barbuy & Grenon, 1991; Barbuy, Ortolani & Bica 1992; Rich 1992). However, very recently McWilliam & Rich (1993) made the first detailed abundance study of Galactic bulge K giants. From their analysis of $\alpha$-elements and iron they concluded that Mg/Fe and Ti/Fe abundances are enhanced by $\simeq 0.3$ dex relative to the solar value over almost the full Fe range, in very good agreement with the predictions of Matteucci & Brocato (1990). On the other hand, they found that two other $\alpha$-elements, Ca and Si, closely follow the normal trend of disk giants. This information looks rather puzzling, although the explanation could be found in a different type of nucleosynthesis and/or a major contribution from type Ia SNe to different $\alpha$-elements. In any case, more observational and theoretical work seems to be required.

For bright elliptical galaxies there are also indications that the $\alpha$-elements are enhanced with respect to Fe. Although, as stated before, determining absolute metallicities in these systems is very hard, it is much better feasible to determine abundance ratios of 2 elements by comparing the strengths of 2 lines of different elements in the same region of the spectrum. Recently this method has been applied to two of the indices of the Burstein system (Burstein et al. 1984), $Mg_2$ and $<Fe>$, by Peletier (1989) and Worthey, Faber & Gonzalez (1992). The reason these indices are used is that they are situated in wavelength regions very close to each other, and are predominantly produced by K-stars. For stars with solar abundance ratios the ratio of $Mg_2$ to $<Fe>$ is more or less constant, except for very cool stars (e.g. Faber et al. 1985, Jacoby, Hunter & Christian 1984). This means that a stellar system composed of stars of solar composition can only occupy a very restricted region in the $Mg_2 - <Fe>$ diagram. If a galaxy is observed outside this range, its abundance ratio of Mg to Fe has to be different from solar. Both Peletier (1989) and Worthey et al. (1992) found that Mg was probably enhanced with respect to Fe in bright elliptical galaxies. However, since at these times no models of stellar evolution with non-solar abundance ratios existed, they could not say by how much.

The aim of this paper is to go beyond this lowest order result about the $Mg_2 - <Fe>$ line strengths. That is, we calculate theoretical line strengths for a variety of parameters of the stellar component of ellipticals. These parameters are (in order of increasing importance) age, mixing length parameter, metal distribution and metallicity. This requires three steps: first, one needs a large set of stellar evolutionary models of various compositions; second, one has to develop a method to calculate line strengths using a given stellar library and making assumptions about age, initial mass functions and missing stellar evolution phases such as the Horizontal Branch; and finally, using the stellar library one has to calculate the theoretical line strengths, compare them with the observations, and draw conclusions about the metal content and thus the evolution of elliptical galaxies.

To begin with the first step, obviously there is a need for evolutionary tracks of super metal rich (SMR) stars with non-solar abundance ratios. While there exist both SMR models with solar metal ratios (e.g. Green, Demarque & King 1987) and metal poor models with $\alpha$-enhancement (e.g. Salaris, Chieffi & Straniero 1993), the combination of both is missing as yet in the literature. To compare theoretical line strengths with observations of ellipticals, we therefore first had to prepare our own library of appropriate stellar models. They will be presented in the first part of this paper in section 2. In the second part, beginning with section 3, the tracks are used to calculate synthetic $Mg_2$ and $<Fe>$ line indices, in the way described in Peletier (1989), and similarly by Gorgas et al. (1993). These indices are then compared with real observations of elliptical galaxies (section 4). On the basis of this comparison we also discuss the effects of various physical parameters on the metallicity indicators in this section, before our conclusions are summarized in section 5.

## 2 Calculation of stellar models

The basic ingredient for any isochrone, line strength, integrated colour, population synthesis and the like are appropriate stellar models. While for globular clusters a large number of calculations and stellar model libraries exist (e.g. Mengel et al. 1979; VandenBerg & Bell 1985; Bell & VandenBerg 1987; Chieffi & Straniero; 1989), this is not the case for the stellar population of ellipticals. The obvious reason lies in the fact that traditionally for the calculation of theoretical models mostly galactic stars have been considered. This implies the combination of high mass with solar or higher metallicities (disk population) or of low mass with very low metallicity (globular clusters in the halo or disk). In the introduction, however, it was argued that there exist good reasons to infer that in ellipticals and probably galactic bulges we might find an old population, i.e. predominantly low mass stars, with nevertheless high metallicities. In particular, if elliptical galaxies are older than $10^{10}$ yrs this implies masses below $1M_\odot$. Integrated colours of elliptical galaxies are so red, and absorption lines, also of our Bulge, so strong, that their metallicities have to be solar or larger, and that models for galactic globular clusters cannot be used.

Consequently, in addition to supersolar metallicity, the effect of relative enhancements of the so-called "$\alpha$-elements" has to be taken into account. Evolutionary tracks for SMR stars with *solar* abundance ratios have already been calculated by various groups, of which we mention here VandenBerg & Laskarides (1987), Green et al. (1987, "Yale") and Charbonnel et al. (1993). Non-solar ratios have only been adopted in stellar evolution calculations aimed at modeling metal poor stars up to now. VandenBerg (1985) already showed that, by assuming [CNO/Fe]> 0, the Main Sequence and Red Giant Branch positions are left unchanged, whereas the turn-off point moves towards a fainter and cooler position, so predicting younger ages for globular clusters than in the standard case with solar abundance ratios. Recently, VandenBerg (1992) has calculated models for oxygen-enhanced compositions (note: not $\alpha$-enhanced), but only for an iron content up to [Fe/H] = $-0.47$ corresponding to a total metallicity Z=0.008.

However, Tornambè (1987) pointed out that observations of stars in the solar neighbourhood and in globular clusters (see Wheeler et al. 1989 for a review) show that, while C and N should not be enhanced, all the $\alpha$-elements (not only oxygen) should be enhanced, and later Chieffi, Straniero & Salaris (1991) argued that different results have to be expected relative to the case where only oxygen, among the $\alpha$-elements, is enhanced. Investigations about the importance of $\alpha$-element-enhancement for low metal stars (Chieffi et al. 1991; Salaris et al. 1993) have shown that the evolution of stars is sufficiently affected to have all the $\alpha$-elements included in the calculations. On the other hand the stellar ages derived in this way do not differ by more than 0.7 Gyr compared to the case where only oxygen or no $\alpha$-element is enhanced at all. Salaris et al. also found that it is only the total metallicity Z that influences the evolution, while the distribution of the elements within Z does not affect it! This result has been doubted by VandenBerg (1992) and remained to be tested for SMR stars. One of the purposes of this paper is therefore to include $\alpha$-element enhancements in the calculations of SMR stars for the first time and to test the effect on the evolutionary tracks.

In this section, we thus present the method of computation and some results of our calculations, in which we have established a data base of stellar models for low-mass stars with solar and above-solar metallicities as well as solar and $\alpha$-enhanced metal ratios. The stellar models were then used for the construction of isochrones and integrated line strengths that will be presented in the next section. A detailed description of the method and all results exceeds the scope of this work, and will be published in a separate paper.

## 2.1 Method of calculation

The stellar evolution program used for these calculations is the one of Weiss (1989), which is based on the classical Kippenhahn code (Kippenhahn, Weigert & Hofmeister 1967). Modifications that have been added later and are significant for this work will be mentioned below as well as relevant details. We start the description with some global assumptions. We have set the mixing length parameter $\alpha_{\mathrm{MLT}}$ to a standard value of 1.5 unless mentioned otherwise. Mass loss was always ignored, since it is mostly unimportant for low-mass main sequence and red giant stars. Also, overshooting from convective cores and undershooting from convective envelopes have not been considered due to the uncertainties in any treatment of these effects and their expected unimportance. A technical detail that turned out to be quite important concerns the evolution along the red giant branch. As is well known in stellar evolution, the computational speed of evolution of low-mass red giants is determined by the pressure gradient above the hydrogen burning shell. This gradient slowly sweeps through the giant's envelope resulting in rather large pressure changes at a few grid points that determine the timestep allowed. At all other grid points, changes are almost negligible. The result is that the evolutionary timestep is of the order of $10^4$ yrs, while the nuclear timescale that determines the evolution physically is ten thousand times larger. Therefore, many programs invoke a shifting of the hydrogen shell on nuclear timescales that rests on certain assumptions about speed of evolution and resulting pressure and temperature profiles (e.g. Sweigart & Gross 1978). Such a description has been used for the Yale isochrones (Green et al. 1987). In our code, however, we followed the evolution exactly, and certain differences in the evolutionary speed of red giants became evident in the course of this project, although they are not critical. As a consequence of our treatment, each evolutionary track from the main sequence to $\log L/L_{\odot} \approx 2.3$ needed 8000 – 10000 time steps, and up to the helium flash about twice as much. We calculated over 250 tracks in total.

Of particular interest for the metal line strengths is a detailed consideration of the metal abundances. This concerns naturally the specification of the abundances in the initial model, but also equation of state, opacities, nuclear energy production, and chemical evolution. The equation of state is that of Weiss (1987), which now considers hydrogen, helium, carbon, nitrogen and oxygen explicitly, but all other elements in a "mean metal" only.

For the nuclear energy production and the chemical evolution we have used a nuclear reaction network for hydrogen burning (see, e.g. Clayton 1983, p. 394). Beta decays are assumed to be in equilibrium; the Ne–Na and Mg–Al cycles (cf. Harris et al. 1983) and the CNO-tricycle $^{17}$O – $^{18}$F were ignored. The reaction rates were those of Caughlan & Fowler (1988). The chemical changes due to nuclear burning were followed by a backward differencing scheme (Arnett & Truran, 1969) between two consecutive stellar models, where the nuclear timestep was controlled by the allowed isotope changes. For small evolutionary timesteps and all cycles being in equilibrium, about ten nuclear steps per evolutionary step were needed.

Our main effort was put into the treatment of opacities. Most of the calculations have been performed before the new OPAL tables of Rogers and Iglesias (1992) became available and rely on the Astrophysical Opacity Library (Hübner et al. 1977; AOL) and the collection of Los Alamos tables of Weiss, Keady & Magee (1990; WKM). We emphasize that the effect of non-solar metal ratios can only be taken into account in the case of the AOL, which allows to compute any mixture of 20 elements. (Some calculations with the new OPAL opacities for solar metal ratios have been added later for comparison; see below.) Our treatment is the following:

We take into account the following six elements: hydrogen, helium, carbon, nitrogen, oxygen, and all other "metals" as the last one. Using six appropriate and linear independent opacity tables, we then solve a set of six linear equations for the weights of the individual tables according to the table compositions. For the abundances we chose relative number fractions to give the dominant element hydrogen a higher weight. Several checks using mass fractions however resulted in negligible changes in the models. Which particular opacity tables we used depended on the composition of the models, but the general scheme was always the same: two tables with a metal abundance and metal ratios close to the initial values (one hydrogen-rich, one hydrogen-free); two tables for typical hydrogen-burning composition (i.e., reduced hydrogen and carbon abundance, increased nitrogen); and finally two standard tables for envelopes and helium cores (WKM1 and WKM21 of Weiss et al. 1990). Table 1 lists the opacity tables and their compositions. All WKM tables were used as published (except for the necessary extension by electron conduction; see Weiss 1987). All other tables were extracted from the Astrophysical Opacity Library for $T > 1$ eV. For lower temperatures they had to be constructed from WKM tables by inter- and extrapolation. Since this involves an additional degree of uncertainty, the interpolation in composition described at the beginning of this paragraph was done only in hydrogen, helium and metals alone for temperatures below 1 eV.

This procedure does ensure that – wherever possible – appropriate opacity tables were used that included effects like $\alpha$-element enhancement or abundances variations due to hydrogen burning, even in the CNO elements. However, for the outermost layers of a star cooler than 1 eV (11605 K) only approximate opacities could be used, since published data for such mixtures (metal rich and $\alpha$-element enhanced) are not available.

**Table 1.** Opacity tables used for the stellar evolution calculations. All numbers are relative mass fractions.

| Table name | source | X | Y | C | N | O |
|---|---|---|---|---|---|---|
| ra1 | WKM 16 | .7000 | .2800 | .0044 | .0011 | .0096 |
| ra2 | WKM 17 | .0000 | .9800 | .0044 | .0011 | .0096 |
| cd2 | WKM 1 | .6020 | .3540 | .0058 | .0019 | .0173 |
| wei2 | WKM 21 | .0000 | .9030 | .0811 | .0011 | .0050 |
| n1_1 | AOL | .7200 | .2600 | .0004 | .0131 | .0016 |
| n1_3 | AOL | .0500 | .9000 | .0009 | .0327 | .0041 |
| a0_1 | AOL | .6500 | .2800 | .0153 | .0037 | .0337 |
| a0_2 | AOL | .1000 | .8400 | .0131 | .0032 | .0289 |
| a0_3 | AOL | .0000 | .9300 | .0153 | .0037 | .0337 |
| a1_1 | AOL | .6500 | .2800 | .0073 | .0018 | .0449 |
| a1_4 | AOL | .0000 | .9300 | .0073 | .0018 | .0449 |
| a2_1 | AOL | .6600 | .2800 | .0045 | .0011 | .0414 |
| a2_3 | AOL | .1000 | .8400 | .0045 | .0011 | .0414 |
| a3_1 | AOL | .6800 | .2800 | .0041 | .0010 | .0257 |
| a3_2 | AOL | .6400 | .3200 | .0042 | .0010 | .0257 |
| a3_3 | AOL | .1000 | .8600 | .0041 | .0010 | .0257 |

The influence of this uncertainty on the models will be seen below, and cannot be avoided. We think, nevertheless, that our treatment of opacity tables is the most advanced one published so far.

## 2.2 Calculations

To cover a broad range in composition that should encompass that of stars in elliptical galaxies, we defined the eight stellar mixtures listed in Table 2. While mixture 0 is basically the solar one, mixtures 1, 2 and 3 have solar metal ratios but higher metallicity $Z$ or varying helium contents. We assumed a $\Delta Y/\Delta Z$ between 0.5 and 2, respectively. Mixtures 5 and 6 correspond to 2 and 3, but with $\alpha$-enhanced metal ratios (see Table 3), and mixture 4 to an intermediate one between 0 and 1. Mixture 7 is similar to 6, but with a slightly higher $X$ and even more enhanced $\alpha$-elements, and therefore corresponds to an earlier galacto-chemical time. Note that Z denotes the *total* metallicity. Due to the enhancement of $\alpha$-elements iron is considerably reduced compared to mixtures of solar metal ratios of the same Z. The total iron abundance in mixtures 4 and 5 is only 0.95 of that in the solar mixture 0. In mixture 7 it is 1.05 Fe$_\odot$. These mixtures would therefore be designated as "solar" on the basis of iron alone. Only mixture 6 (1.66 Fe$_\odot$) is super-solar in all respects.

We need to mention here that the correlation between helium and metals is a major uncertainty. Our values ranging from 0.5 to 2 are very conservative. In fact, there is still debate on this value. On the one hand, Pagel (1992) indicates values between 4 and 5 from observations of extragalactic HII regions. On the other hand, an estimate of this ratio can be made just by considering a primordial He abundance of 0.24 and the solar He abundance of 0.28. The derived ratio is in this case exactly 2, when a solar metallicity of 0.02 is considered. From the theoretical point of view, standard chemical evolution models generally imply values between 1 and 2, except with Maeder's (1992) new yields, which predict a $\Delta Y/\Delta Z$ in agreement with Pagel's estimate. In this work it is assumed that stars above $20 - 25 M_\odot$ form black holes and consequently do not contribute to the metal enrichment of the galaxy.

**Table 2.** Initial chemical composition (mass fractions) of stellar models. Z is the total metallicity including C, N and O.

| mixture | X | Y | C | N | O | Z | type |
|---------|-----|-----|-------|-------|-------|------|------------|
| 0 | .70 | .28 | .0044 | .0011 | .0096 | 0.02 | solar |
| 1 | .70 | .26 | .0087 | .0021 | .0193 | 0.04 | solar |
| 2 | .64 | .32 | .0087 | .0021 | .0193 | 0.04 | solar |
| 3 | .65 | .28 | .0152 | .0037 | .0337 | 0.07 | solar |
| 4 | .68 | .28 | .0041 | .0010 | .0257 | 0.04 | $\alpha_1$ |
| 5 | .64 | .32 | .0041 | .0010 | .0257 | 0.04 | $\alpha_1$ |
| 6 | .65 | .28 | .0073 | .0018 | .0449 | 0.07 | $\alpha_1$ |
| 7 | .66 | .28 | .0045 | .0011 | .0414 | 0.06 | $\alpha_2$ |

The two $\alpha$-enhanced mixtures, labeled $\alpha_1$ and $\alpha_2$, are listed in Table 3 and follow from predictions of chemical evolution models. In particular, the enrichment in oxygen is [O/Fe]=0.45 and 0.67, respectively (Matteucci 1992a), since these values are typical for halo stars (Wheeler et al. 1989). The enrichment is not the same for all $\alpha$-elements, but very similar.

**Table 3.** Relative element abundances (mass fractions) in metal types.

| element | solar | $\alpha_1$ | $\alpha_2$ |
|---------|---------|---------|---------|
| C | 0.21785 | 0.10366 | 0.07434 |
| N | 0.05308 | 0.02526 | 0.01811 |
| O | 0.48158 | 0.64161 | 0.69021 |
| Ne | 0.03262 | 0.04346 | 0.04675 |
| Na | 0.00190 | 0.00090 | 0.00065 |
| Mg | 0.04211 | 0.05009 | 0.05748 |
| Al | 0.00392 | 0.00186 | 0.00133 |
| Si | 0.05458 | 0.05843 | 0.05122 |
| P | 0.00043 | 0.00020 | 0.00015 |
| S | 0.02208 | 0.02944 | 0.02677 |
| Cl | 0.00049 | 0.00023 | 0.00017 |
| Ar | 0.00175 | 0.00083 | 0.00060 |
| Ca | 0.00391 | 0.00419 | 0.00367 |
| Ti | 0.00023 | 0.00011 | 0.00008 |
| Cr | 0.00115 | 0.00055 | 0.00039 |
| Mn | 0.00063 | 0.00030 | 0.00021 |
| Fe | 0.07684 | 0.03656 | 0.02622 |
| Ni | 0.00485 | 0.00231 | 0.00166 |

For each composition we calculated the following 17 masses (in units of $M_\odot$): 1.10, 1.05, 1.00, 0.98, 0.96, ..., 0.80, 0.75, ..., 0.60. We started with homogeneous models on the ZAMS. This implies that the CN-subcycle first had to come into equilibrium. The resulting loop in the Hertzsprung-Russell-Diagram (HRD) lasted several $10^7$ yrs and is omitted in the figures (see Mazzitelli & D'Antona 1986 for the same procedure). We followed the evolution on the Red Giant Branch until a luminosity of approximately $\log L/L_\odot = 2.3$ or an age of $25 \cdot 10^9$ yrs was reached. Some tracks were also carried into the core helium flash to allow an extension of all other giant branches. Horizontal Branch and Asymptotic Giant Branch stars were not considered. The necessary corrections to the computed metal line strengths will be discussed in the following sections.

## 2.3 Results of the stellar evolution calculations

For the calculation of metal line strengths luminosity, effective temperature and luminosity function (i.e. evolutionary speed) are of major importance besides the surface abundances of the elements. In our discussion of the evolutionary results we will therefore concentrate on these quantities. Other properties of the evolution will be discussed in a more extended paper.

Fig. 1 shows as a typical example the evolutionary tracks in the HRD for selected masses calculated in the case of mixture 2. Two of the tracks were followed into the core helium flash to provide a basis for an extrapolation of all the other giant branches. This approximation saves more than 50% of the computations. The higher density of stars around $\log L/L_\odot \approx 1.2$ ("clump stars") is evident. The first dredge-up results in a slight increase in the surface helium abundance of 0.01 to 0.02. Also, carbon becomes slightly reduced as compared to the initial value due to CN-burning and dredge-up. Mg and Fe remain unchanged, since in our calculations there exists no physical process to alter their abundances; changes in oxygen are negligible.

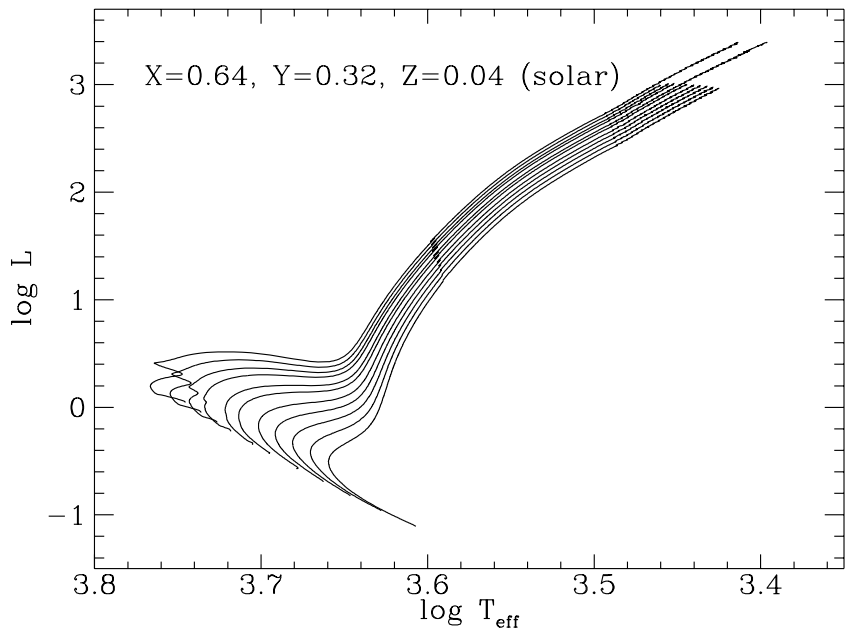

**Fig. 1.** Evolution of stars of composition 2 from the ZAMS to the end of the calculations. For clarity several masses have been left out. The models of 1.1 and 0.9 $M_\odot$ were followed into the core helium flash.

The influence on the HRD of changing the metal content from 0.02 to 0.07 is shown in Fig. 2 for a stellar mass of $0.9 M_\odot$; that of changing the helium content (or, to be precise, the assumption about $\Delta Y/\Delta Z$) in Fig. 3 ($Z$=0.04; $\alpha_1$), and that of changing the metal abundance ratios in Fig. 4. The changes in the evolutionary speed are not shown. Since they enter the stellar constituents in a population of given age, they will be discussed in section 4 in the context of the influence of age on the line strengths.

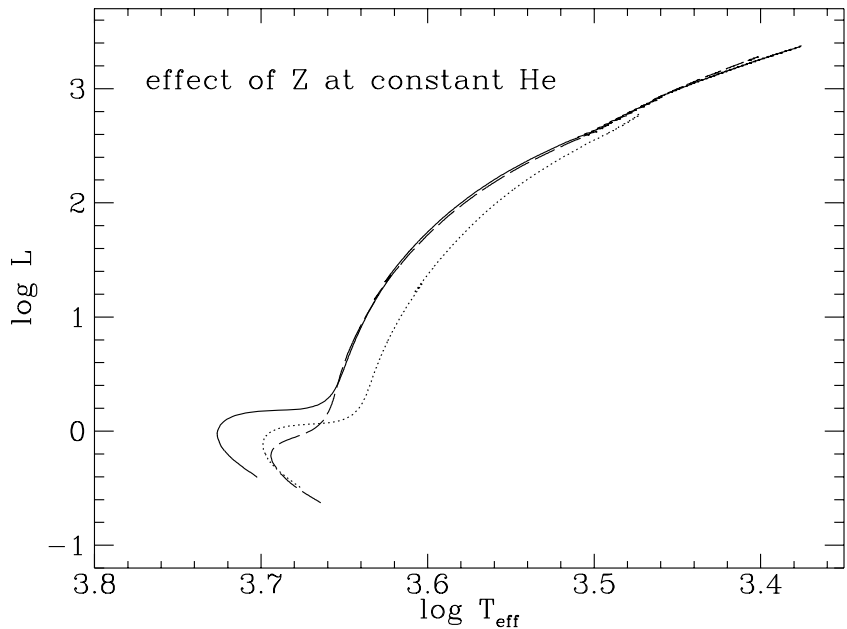

**Fig. 2.** Effect of increasing the total metallicity in the case of a model of 0.9 $M_\odot$ and mixtures 0 (solid line), 1 (dotted) and 3 (dashed)

From Figs. 2–4 it is apparent that the biggest changes due to variations in the compositions are on the main sequence. The effective temperatures along the red giant branch agree within 5% or less. At least partially, this might be due to unreflected composition dependencies in the opacity tables at lowest temperatures (see 2.1). The tendency is that higher metallicity or lower helium content leads to lower main sequence luminosities and effective temperatures. The same happens by the transition of solar to $\alpha$-enriched metals in the case of mixtures 2 and 5 (Fig. 4; $Z = 0.04$). At the same time the red giant temperatures become higher. The same changes, but on a much smaller level take place in the case going from mixture 3 to 6 ($Z = 0.07$; solar to $\alpha_1$).

One particularity seen in Fig. 2 should be mentioned: the helium content of mixture 1 (dotted line) is lower by 0.02 than that of the two other mixtures compared. From Fig. 3 we deduce that the main sequence part of this track should be shifted by $\approx +0.015$ in $\log T_{\rm eff}$ to correspond to a mixture with Y=0.28. Then temperature and luminosity for a metallicity of 0.07 (dashed line) are below the less metal rich models, as should be expected from the comparison of Z=0.02 and 0.04. After the turn-off, however, the evolution approaches more and more that of the solar metallicity case, i.e. temperatures are higher than for Z=0.04. This unexpected and unsystematic behaviour could be due to the fact that for Z=0.07 the opacities, which determine the temperature gradient in the outermost layers, had to be extrapolated from tables with lower metal content for $T \leq 1\,\mathrm{eV}$, and therefore are not very reliable. The consequences of this will be seen later on in section 3.4 (Fig. 7).

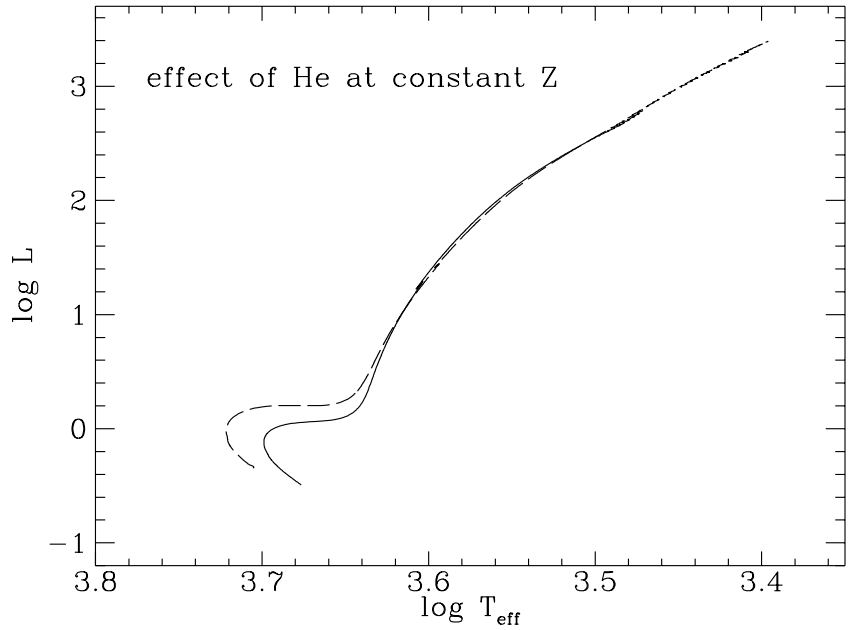

**Fig. 3.** Effect of changing helium at constant metallicity (0.04; $\alpha_1$) in the case of a model of 0.9 $M_\odot$ and mixtures 4 (solid) and 5 (dashed)

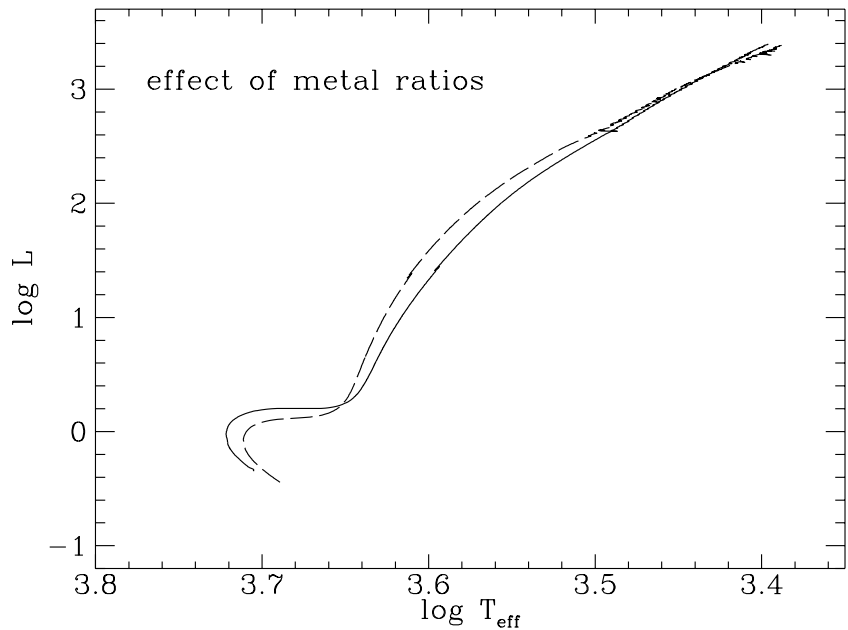

**Fig. 4.** Effect of changing the metal ratios from solar to $\alpha_1$ in the case of a model of 0.9 $M_\odot$ and mixtures 2 (solid) and 5 (dashed)

While we have decided to do comparisons between solar and $\alpha$-enhanced mixtures on the basis of constant total metallicity Z (the theorist's metallicity), one could also do comparisons for a constant iron abundance (the observer's metallicity). In this sense, our mixtures 0 and 7 have (almost) the same iron content ($1.57\,10^{-3}$ and $1.54\,10^{-3}$), and the higher metallicity Z is due to the $\alpha$-enhancement only. It is important to recall that this is also the approach of Salaris et al. (1993): their metallicity Z corresponds *only* to that mass fraction of metals that one obtains by scaling solar abundances to a given iron content. In the case of $\alpha$-element enhancement this implies that there is an *additional* metal fraction present. In particular, for an enhancement of $\alpha$-elements by a factor of 4 this leads to a *total* or *global* (their notation) metallicity almost three times the value of Z as used by Salaris et al! Therefore, if they conclude that $\alpha$-element enhancements have to be taken into account in the calculations they actually mean that the total metal content must be used and not just a metallicity obtained by scaling a model with solar composition to the observed iron abundance, because this value might be much too low!

One of the surprising findings of Salaris et al. is that once the global metallicity is used correctly, the internal distribution of elements within the metal group does not affect the evolution, implying that solar metal ratios can be used if they are only scaled to the global metallicity. This result has been doubted by VandenBerg (1992) for metal-poor stars. From our Fig. 4, which compares two evolutionary tracks at the same global metallicity but different internal element ratios, we conclude that in our calculation of SMR stars the individual abundances do affect the evolution clearly, though much less than the total metallicity. Therefore, we found it worthwhile and necessary to do the $\alpha$-enhanced calculations for supersolar metallicity explicitly. This contradiction to Salaris et al. (1993) might, however, be due to the fact that the quantity $[(X_C + X_N + X_O + X_{Ne})/(X_{Mg} + X_{Si} + X_S + X_{Ca} + X_{Fe})]$ defined in Salaris et al. (1993) is approximately 0.10 for our $\alpha_1$ mixtures. Salaris et al. (1993) argued that it should be 0 for two evolutionary tracks at the same global Z to agree even though the internal metal ratios are different.

In addition to our standard choice of parameters we added calculations with a mixing length parameter $\alpha_{MLT} = 2.5$ for mixtures 0, 2, 5 and 7. The resulting metal line strengths will be shown in section 4, but we refrain from showing the evolutionary changes brought about by this variation. Roughly, effective temperatures increase by $+0.05$ dex for all mixtures and evolutionary phases. These calculations were needed to study the influence of the uncertainties in the theory of convection and to allow a calibration of $\alpha_{MLT}$ (section 3).

We finally add some remarks on the use of the opacity tables. The tables collected by Weiss et al. (1990) fall into two distinct groups: those that level off to $\log\kappa \approx -2.5$ and those that drop to $\log\kappa \approx -5.5$ at the lowest temperatures. While in all calculations cited up to now we used tables of the first group, we added one run (mixture 2) with table WKM14 (a table of the latter group) for the hydrogen-rich envelope. The resulting evolutionary tracks were hotter by $\Delta\log T_{eff} \approx 0.1$ (!) both for the main sequence and the red giant branch, and were disregarded therefore. Of course, such a change could be compensated by an adjusted $\alpha_{MLT}$, but then the relative differences of the calculations could not be connected with the mixtures alone, and such our study would be worthless. This demonstrates once again the need for consistent opacities tables.

After most of the calculations were completed, the first opacity tables of the Livermore group (Rogers & Iglesias 1992) became available. They all have solar metal abundances, but various hydrogen and helium contents. In order to test their influence on our results, we implemented them with a quite different opacity routine package (details will be given elsewhere). The evolutionary tracks changed in many respects (main sequence luminosity and temperature; turn-off luminosity; RGB temperature), but these changes were always very small (e.g. $|\delta T_{\text{eff}}| < 0.02$) with different signs for the chemical compositions investigated. The resulting change in the metal line strengths will be shown in section 4. We therefore conclude that the calculations we presented above are still representative and do not have to be replaced up to now, in particular because of the fact that $\alpha$-element variations could not be accounted for with the OPAL tables.

## 2.4 Comparison with earlier calculations

In the course of this work it became evident that our results for the metal line strengths differ from those obtained using the revised Yale isochrones (Green et al. 1987). These differences will be discussed in section 4, but they result from the underlying stellar models. Since the Yale isochrones rely on stellar evolution calculations almost 15 years old, some differences should be expected, naturally. However, we decided to check our stellar evolution program with more recent publications that use similar physics and possibly codes.

The most recent and complete set of stellar models for solar metallicity and a wide mass range has been published by Maeder & Meynet (1988). We thus compared their $1M_{\odot}$ star ($\alpha_{\text{MLT}} = 1.9$; $X = 0.70$, $Z = 0.02$) with a model calculated for the same set of parameters by our program. Maeder's code has evolved from the same original Kippenhahn program, uses the latest Los Alamos Opacities as well, and should therefore compare best with our one. The inclusion of overshooting and mass loss in Maeder & Meynet (1988) is not expected to lead to any big difference for a $1M_{\odot}$ star. Indeed, for a model on the main sequence (age $3.25\,10^9$ yrs) we find an excess in the luminosity of our model of only $\Delta \log L/L_{\odot} = 0.003$ and a temperature increase of $\Delta \log T_{\text{eff}} = 0.0015$. Along the Red Giant Branch, the effective temperature agrees with that of Maeder & Meynet at any given luminosity within 0.003. Stellar ages agree better than 2% on the main sequence and 5% along the RGB.

In a second study we compared with results of VandenBerg (1983) for the parameters $M = 1M_{\odot}$, $\alpha_{\text{MLT}} = 1.0$ and $1.3$, $Y = 0.30$, $Z = 0.0169$. Differences here are about five times larger than in the first case, but still the agreement is very good. In particular the effective temperature of the Red Giant Branch agrees within less than $150\,\text{K}$ and main sequence lifetimes differ by $5\,10^8$ yrs at most. We thus conclude that our code can reproduce comparable calculations and the results can safely be used as input for the calculation of metal line strengths.

As a next step we localized the differences in the computed metal line strengths (see section 4) between our calculations and those employing the Yale isochrones (for the same set of parameters) to a problem on the lower main sequence, which is presented in Fig. 5. Here, mass versus effective temperature is shown for a given age (10 Gyr).

While the turn-off mass for the Yale models and ours with similar composition (Y=0.26; long-dashed) agrees very well, the slope of the Yale main-sequence line is much shallower. This results in a higher contribution to the integrated light from the lower main sequence and is equivalent to having more stars of low $Mg_2/<Fe>$ so that the integrated value of $Mg_2/<Fe>$ will be somewhat lower for the Yale models, too (see section 4). Our second composition shown in Fig. 5 (Y=0.32) agrees very well with the first one with respect to the slope, but has a lower turn-off mass, as should be expected from the fact that both the hydrogen content is lower (less fuel), but the luminosities are higher (higher molecular weight).

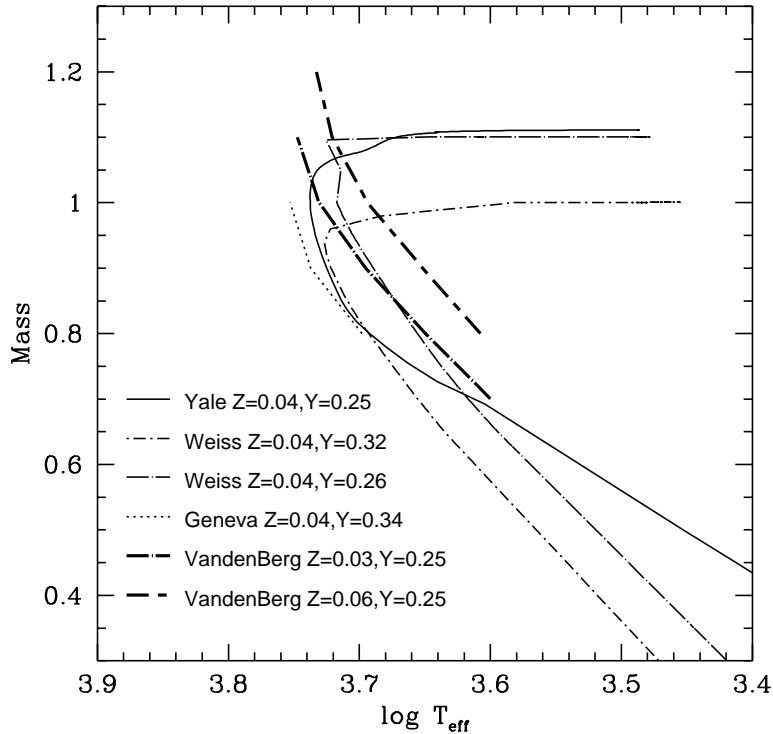

**Fig. 5.** Mass vs. effective temperature for various authors

In addition to these models, we show the equivalent tracks from VandenBerg & Laskarides (1987) and Charbonnel et al. (1993) (Geneva). Again, they agree with our results, but differ from Yale. Not shown in Fig. 5 are data from Maeder & Meynet (1988) and VandenBerg (1983) for solar mixtures, different ages, and various $\alpha_{MLT}$ thay we have checked as well. They all are in agreement with respect to the lower main sequence slope. It appears that the structure of the lower main sequence stars differs appreciably between the Yale models and those with more updated physics, and leads – via different colour contribution to the integrated light – to the modest line strength differences that will be discussed in the next section. We emphasize that the differences are not between Yale and this work, but between Yale and (most likely) all recent stellar evolution calculations.

## 3 Synthesizing integrated indices

In this paper an approach has been followed similar to the one described in Peletier, Valentijn & Jameson (1990) (PVJ) and by Gorgas et al. (1993). The parameters of a star in the theoretical HR diagram, temperature, gravity and luminosity, are transformed to observable parameters, after which we choose a mass function, and determine indices for single age stellar populations by integrating along the HR diagram. For optical and infrared broadband colours, the procedure has been described in PVJ, while for line indices it can be found in Peletier (1989). We will repeat here the essential steps that were employed.

## 3.1 From the theoretical to the observational HR diagram

To calculate stellar colours from the models, we first converted the effective temperatures to $V-K$, and then converted $V-K$ to other colours, like $U-V$, $B-V$ and $V-I$ using observed dwarfs and giants. This first step is accurate, because $V-K$ is a good temperature indicator, since line blanketing is not very important here. The $V-K$ – temperature calibration comes from Ridgway et al. (1980), who could measure stellar radii for giants and $V-K$ using a linear occultation technique, and as such also the effective temperature. For the conversion from $V-K$ to other colours we use observed stars from Johnson (1966), Frogel et al. (1978) and Allen (1973). For the colour-colour conversion a distinction was made between dwarfs and giants. Since these stars all have a composition similar to solar, the integrated colours will only be valid for metallicities not very different from solar.

For line strengths, the previous approach does not work. Due to the extreme importance of line blanketing on the indices, the conversion between $V-K$ and e.g. $Mg_2$ is different for each metallicity. To obtain the $Mg_2$ index for a given Mg abundance and a certain temperature (as given by $V-K$) in the ideal case one needs a library of stars of various metallicities for which the Mg abundance, as well as the $V-K$ colour, has been determined. Unfortunately, this is not available, but the library by Rich (1988) of galactic bulge stars is a good alternative. It contains stars of high metallicity, for which individual metallicities [Fe/H] were determined in a variety of ways, described in that paper. A least squares fit to these stars gave:

$$
\begin{aligned}
[\text{Fe/H}] = \ & 2.247 - 1.615\,(V-K) + 2.518\,Mg_2 \\
& + 1.814\,(V-K)\,Mg_2 - 4.9128\,Mg_2{}^2 \quad (2 \le (V-K) \le 4)
\end{aligned}
\tag{1}
$$

and

$$
\begin{aligned}
[\text{Fe/H}] = \ & 1.751 - 1.365\,(V-K) - 3.054\,\log(\langle\text{Fe}\rangle) \\
& + 1.987\,(V-K)\,\log(\langle\text{Fe}\rangle) + 1.8345\,\log(\langle\text{Fe}\rangle)^2 \\
& (2 \le (V-K) \le 4)
\end{aligned}
\tag{2}
$$

Since Rich observed stars near the Galactic center, his sample consists of giants of type K or later $(2 \le (V-K) \le 4)$ only; other stars are too faint. For earlier types of stars, the metallicity dependence of the line indices in the SMR region is not well known. This however is not such a big problem for the <Fe> and $Mg_2$ indices, since these are dominated by K stars. For hotter stars the lines are very weak, while for cooler stars the relative contribution in the region of the spectrum around 5000 Å is small. As a first approximation, <Fe> and $Mg_2$ for stars with $(V-K) < 2$ and $(V-K) > 4$ were calculated purely as a function of the temperature indicator, $V-K$. To get the calibration, we measured the $Mg_2$ and <Fe> indices for all Main Sequence and Giant Branch stars from the stellar library of Jacoby et al. (1984) and transformed their stellar $B-R$ colours into $V-K$ (see PVJ). Using least squares fits, the indices are obtained with the following recipe:

$$\text{Mg}_2 = \begin{cases} 0.0265(V{-}K) + 0.0205 & (-1 \le V{-}K \le 1) \\ 0.136(V{-}K) - 0.089 & (1 \le V{-}K \le 2) \\ \text{from (1), but} \le 0.62 & (2 \le V{-}K \le 4) \\ 0.62 & (V{-}K \ge 4) \end{cases} \qquad (3)$$

and

$$\langle \text{Fe} \rangle = \begin{cases} 0.613(V{-}K) + 0.413 & (-1 \le V{-}K \le 0.6) \\ 1.814(V{-}K) - 0.309 & (0.6 \le V{-}K \le 2) \\ \text{from (2)} & (2 \le V{-}K \le 4) \\ -1.291V{-}K + 9.294 & (4 \le V{-}K \le 7.2) \\ 0 & (V{-}K \ge 7.2) \end{cases} \qquad (4)$$

After integration, the errors in the two indices introduced by neglecting the effects of line blanketing for the hot and the very red stars are of the order of a few percent.

## 3.2 Integrating along the HR diagram

The method used here has been described in PVJ. Since evolutionary tracks were only calculated for masses between 0.6 and 1.1 $M_\odot$, parameters for lower masses were determined by extrapolating linearly downward in mass. As an initial mass function we took the Salpeter mass function ($dn/dm \propto m^{-s}$, $s = 2.35$), since the IMF in elliptical galaxies is not very well known, and the results are not very dependent on the IMF (cf. Worthey 1992; Peletier 1989). For the Salpeter mass function the light from stars with masses smaller than 0.6 $M_\odot$ is minor compared to the turnoff region and the Sub Giant Branch.

Since our models were calculated only to the tip of the RGB, we had to find some other way to estimate the contribution to the light from later stages of stellar evolution. In order to avoid introducing free parameters, we used a simple recipe following from the fuel consumption theorem to estimate at each age what the ratio of the light from AGB to RGB was (see Renzini & Fusi-Pecci 1988). (What we call AGB in this paper consists of Red Horizontal Branch and AGB). AGB colours were chosen to be the same as those of the RGB. The effect of introducing the AGB has been analyzed in PVJ and will be again in the next section. No blue Horizontal Branch was included, because it is lacking in globular clusters with metallicities higher than that of 47 Tuc. To convert bolometric luminosities to luminosities in the V-band, we used a recipe from Bessell & Wood (1984), as described in PVJ.

## 3.3 Comparing the isochrones with observations: M 67

A check of the reliability of the isochrones is the comparison with observational data. Since the objects that are studied in this paper are of metallicities larger or equal to solar, we have compared the isochrones to colour-magnitude diagrams of some metal rich objects. In this section a comparison is made with the open cluster M 67, (data from Eggen & Sandage (1964)), while in the next section a comparison is made with three globular clusters. We used our mixture 0, with metallicity Z=0.02, and an age of 4.5 Gyr (Anthony-Twarog 1987). The isochrones have been determined for a mixing length to pressure scale height parameter $\alpha_{\text{MLT}}$ of 1.5 and 2.5. The data were corrected for a foreground absorption of E(B-V) = 0.06 mag, and a distance modulus of 9.38 was used (Eggen & Sandage 1964). Results are given in Fig. 6. The agreement with the data is reasonable, especially since one has to realise that the metallicity used here is higher than the one derived by Anthony-Twarog (1987) ([Fe/H] = 0.20 vs. −0.07), and also since it is well possible that different isochrones give a different age-metallicity combination. As far as $\alpha_{MLT}$ is concerned: it appears that the true value to be used for the giant branch should be around 1.8 - 2. Indeed, a calibration of $\alpha_{MLT}$ with a solar model gives 2.0 as well (see the solar model published in Kippenhahn & Weigert 1989, p. 371).

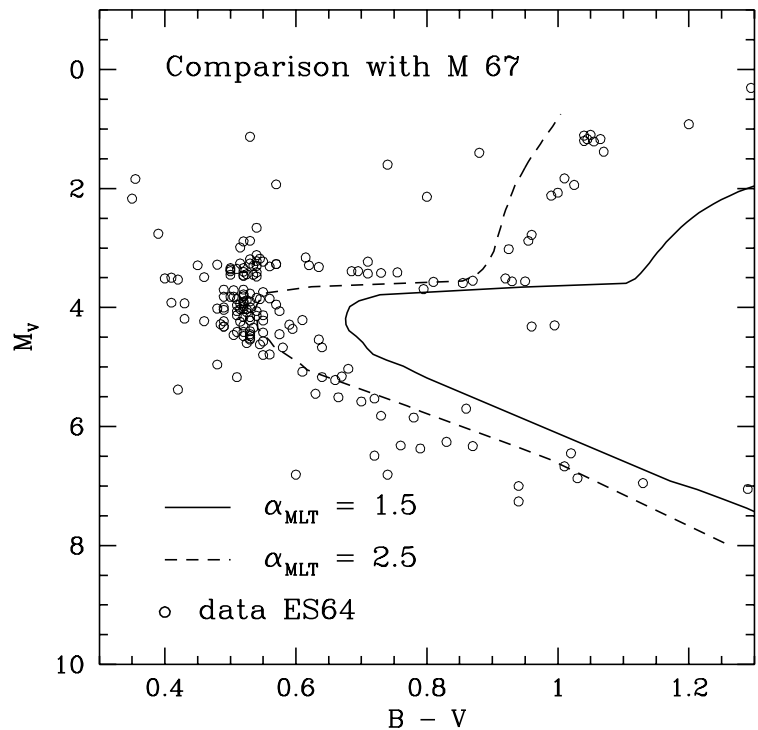

**Fig. 6.** Colour-magnitude diagram of the open cluster M 67 (data from Eggen & Sandage 1964). Superimposed are our isochrones with Z=0.020 and age=4.5 Gyr, with $\alpha_{MLT}$ = 1.5 and 2.5.

### 3.4 Colour-magnitude diagrams of super metal rich globular clusters

Stellar evolutionary models for metallicities larger than solar are difficult to test given the lack of super metal rich objects that can be resolved into stars. There are however some strong-lined globular clusters in the direction of the bulge, whose colour-magnitude diagrams can be used. For testing purposes, these globular clusters unfortunately are not ideal, since they suffer from extinction by dust because of their location. In the near future, the availability of near-infrared photometry will significantly improve this situation. There is also hope that high-resolution observations will be available of the central areas of M 31 and M 32, so that good comparisons with metal rich galaxies are possible.

Not very much work has been done to study globular clusters in the bulge of our Galaxy, probably because their individual stars are faint, and for the reasons mentioned above. In this paper we will compare some of our models with three globular clusters presented by Bica, Barbuy & Ortolani (1991).

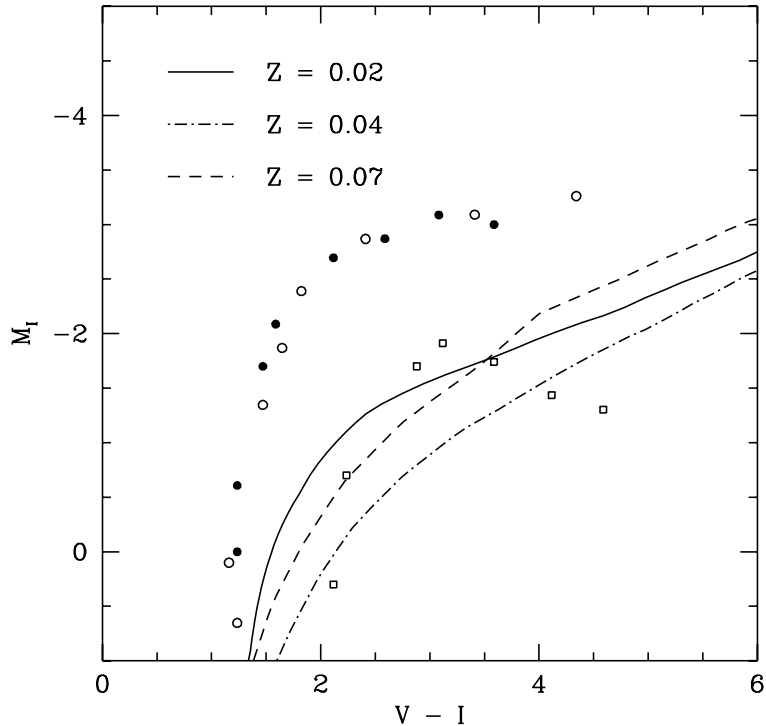

**Fig. 7.** Colour-magnitude diagram of the giant branches of three globular clusters in the Galactic bulge. Plotted are data from Bica et al. (1991). Terzan 1 is indicated with open squares, NGC 6528 with open circles and NGC 6553 with filled circles. Also plotted are three of our models with solar abundance ratios.

In Fig. 7 we present NGC 6528, NGC 6553 and Terzan 1 in the $V-I$ vs. $M_I$ diagram. These globular clusters are more metal rich than 47 Tuc (Bica et al. 1991), which shows up in the fact that the stars are redder, and that the most luminous stars are fainter in $I$. From integrated quantities Zinn & West (1984) found a metallicity [M/H] of −0.29 for NGC 6553. Recently, this was confirmed by a high-resolution study by Barbuy et al. (1992), who found a metallicity of $-0.2^{+0.2}_{-0.4}$. The metallicity of NGC 6528 should be very similar, since it has almost the same colour-luminosity relation. One peculiar aspect of Terzan 1 is that its giant branch is bending over. This effect is usually ascribed to line blanketing, although part of it is because for red stars most of the light has gone into redder bands. The effect has also been seen for other clusters like 47 Tuc (Lloyd Evans & Menzies 1977), so it does not occur only for extremely metal rich globular clusters. For NGC 6528 and NGC 6553 however there does not seem to be any turnover. Even the turnover for Terzan 1 is caused by only a few stars, whose magnitudes could be affected by differential reddening.

We have plotted 3 of our models in the diagram, all for 20 Gyr, and solar metal ratios, but for three different metallicities. The shape of the models at Z=0.04 and Z=0.02 agrees reasonably well with with NGC 6528 and NGC 6553, but both are more than 1.5 mag fainter than the globular clusters. However, since the metallicity of NGC 6553 is probably –0.29, we should have used a metallicity that is lower by a factor 2.3, so that the isochrone will brighten more or less to the correct value. A further problem seen in Fig. 7 is the fact that the tip of the RGB of the models does not bend down. However, since the stars here are too cool to affect significantly the $Mg_2$ and $<Fe>$ indices this does not influence the results of this work.

When going from Z=0.02 to Z=0.04 the giant branch becomes indeed redder and fainter, as one expects if Terzan 1 is more metal rich than NGC 6528 and NGC 6553, but at Z=0.07 the giant branch turns bluer dramatically. This effect is also seen in the Yale isochrones for Z=0.10. In our case this can be traced back to the effective temperature variations shown in Fig. 2, which are non-monotonic as well. The fact that the RGB at Z=0.07 crosses the RGB at Z=0.02 is probably an artifact of the large bolometric correction which are applied for very cool stars and might not be real.

# 4 Results

## 4.1 Old models using Yale tracks with solar abundances

In previous papers, $Mg_2$ and $<Fe>$ were modeled in a similar way as here, with the difference that the stellar models were calculated using solar abundance ratios for all elements except H and He (Peletier 1989, Worthey et al. 1992). It was shown in both papers that for a given value of $Mg_2$ the models produced a $<Fe>$ index that was too large to fit the observations of bright elliptical galaxies, independent of age, metallicity or IMF. If the models are correct, this implies that abundance ratios (e.g. [Mg/Fe]) have to be different from solar. To a first approximation one can assume that the stellar models (i.e. log $T_{eff}$, L and log g) remain the same if Z, the mass fraction of heavy elements, is kept constant, but the ratios of the individual elements are varied. Metal line strengths obtained in this manner from the Yale stellar models are presented in Fig. 8 for solar abundance ratios and for compositions $\alpha_1$ and $\alpha_2$, for Z=0.01 and Z=0.04, and ages 12, 15 and 18 Gyr.

The data in this figure are from Davies, Sadler & Peletier (1993), currently the paper with the highest quality radial profiles of $Mg_2$ and $<Fe>$ for elliptical galaxies. Displayed are radial profiles for 9 ellipticals with $M_B$ between −20 and −23. The dashed line indicates the best fit to data for the nuclei of many elliptical galaxies, from Burstein et al. (1984). These models give the impression that the [Mg/Fe] abundance ratios for these 9 giant elliptical galaxies (small symbols in Fig. 8) lie between [Mg/Fe]=0.45 ($\alpha_1$) and 0.67 ($\alpha_2$), and that [Mg/Fe] as a function of radius in a galaxy remains constant, or increases slightly. With regard to the line from Burstein et al. (1984) these models imply that [Mg/Fe] is larger than solar ($\alpha_1 - \alpha_2$) for bright galaxies, and gets closer to solar for fainter galaxies. All metallicities considered here lie between Z=0.01 and Z=0.04, for ages between 12 and 18 Gyr.

## 4.2 New models

The main results of this paper are presented in Fig. 9 and Table 4. Here we plot the same quantities as in Fig. 8 against each other, but now we present the result for our new models, for single ages of 12, 15 and 18 Gyr. The thin dashed line again indicates the data by Burstein et al. (1984). The thin solid line connects the previous theoretical models for 18 Gyr, that use solar abundance ratios and Yale isochrones.

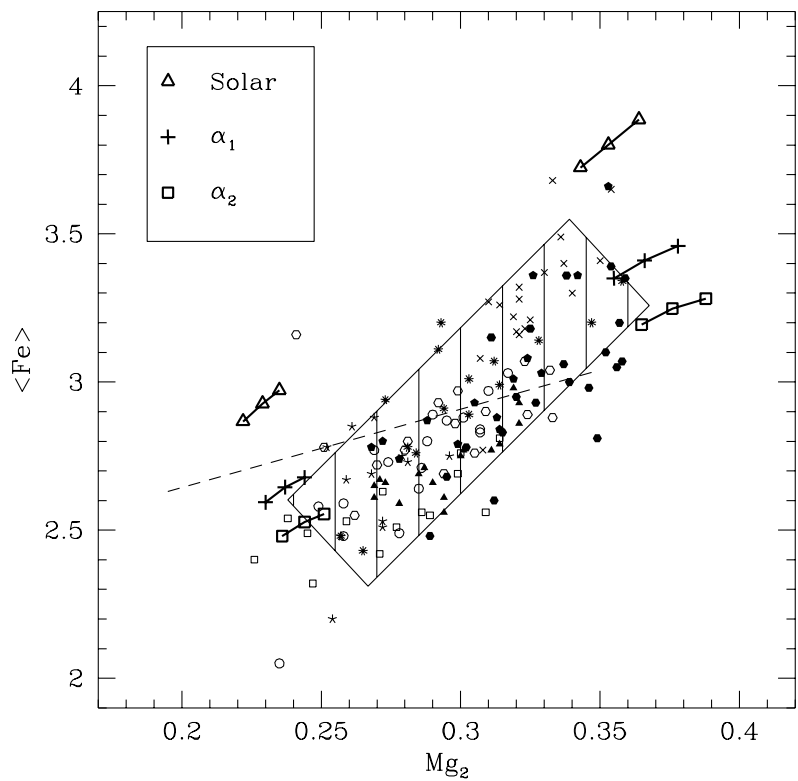

**Fig. 8.** Mg$_2$ and <Fe> for old models using the Yale isochrones. Models are indicated by large symbols and fat lines, for Z=0.01 (left group) and 0.04 (right group) and ages of 12, 15 and 18 Gyr (ages increase from left to right along one line of models). Plotted are the radial profiles of 9 giant ellipticals with the nuclei in the upper right corner. The dashed line indicates the best fit to the data for the nuclei of many elliptical galaxies, from Burstein et al. (1984). The shaded box has been drawn to contain most of the data points.

Comparison of the new models with the old Yale-models shows that for a given metallicity both Mg$_2$ and <Fe> are larger using the new models, but also that <Fe> increases somewhat for a given Mg$_2$ index. The first effect can be explained by the redder turnoff of the new models, caused by the fact that the new opacities are larger, while the second is due to the fact that the mass-temperature relation is different from the old Yale models (Fig. 5). This has been discussed in section 2.4, and has been shown to be a particularity of the Yale models when compared to more recent calculations using updated physics.

The effect of changing [Mg/Fe] with a constant mass fraction of elements heavier than H and He (Z) is mainly that Mg$_2$ remains constant while <Fe> decreases (e.g. from mixture 2 to 5). Since the mass fraction of Fe, as compared to $\alpha$-elements like O, Mg, Na etc. is small, a change in [Mg/Fe] basically means an increase or decrease in Fe, while the Mg abundance remains more or less constant. When the stellar models are not affected much by the abundance ratios, one then expects this behavior in Fig. 9. Model 5H is the same as model 5, except for the fact that here the luminosities and effective temperatures of model 2 are used. The fact that models 5 and 5H fall almost on top of each other shows that the effects of taking into account $\alpha$-enhancment in the stellar models are small compared to all other parameters.

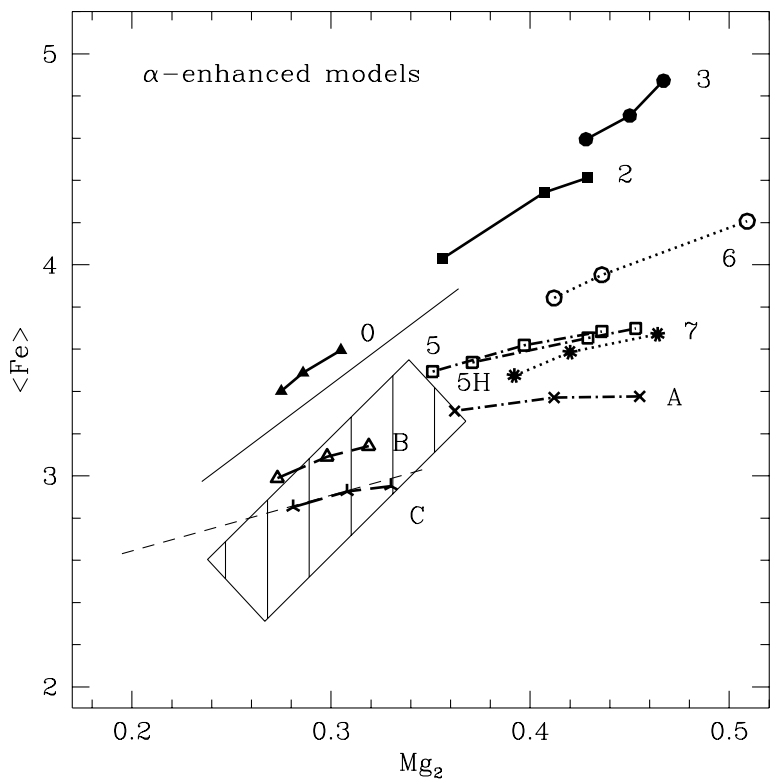

**Fig. 9.** Mg$_2$ and <Fe> for models with $\alpha$-enhanced abundances for ages of 12, 15 and 18 Gyr. Digits indicate the mixtures of Table 2. Models A, B and C are hybrid models - model A uses the stellar interior mixture 4 but has composition $\alpha_2$, while B and C use the interior mixture 0 for composition $\alpha_1$ and $\alpha_2$, resp. The dashed line indicates the mean position of the nuclei of the galaxies of Burstein et al. (1984), the shaded box the position of the data of Davies et al. (1993), and the thin drawn line the position of old models with solar composition for 18 Gyr (section 4.1).

In Fig. 10 we investigate what happens if we add or omit a group of stars that is not described well by the stellar models. The figure shows in the first place that if we do not correct for the contribution from AGB stars, it will barely affect the Mg$_2$ and <Fe> indices. This is because the contribution of AGB stars in the V-band, in which the line indices are situated, is not very large. The change in indices as a result of adding a Blue Horizontal Branch (BHB) is more significant. In Fig. 10 we have indicated for the models of solar composition for Z=0.02 and 0.04 what the effect is of adding a BHB that constitutes 5% of the light in V. For this BHB we have assumed Mg$_2$ =0.047 and <Fe> =1.5. Although one expects a red HB for elliptical galaxies, on the basis of colour-magnitude diagrams of metal rich globular clusters, there are indications from population synthesis that a significant component of blue stars is needed as well. For this

**Table .** Mg$_2$ and <Fe> for the new models

| [Mg/Fe] | Z | Y | $\alpha_{\mathrm{MLT}}$ | Mix | Age [Gyr] | Mg$_2$ | <Fe> |
|---|---|---|---|---|---|---|---|
| 0.00 | 0.02 | 0.28 | 1.5 | 0 | 12 | 0.263 | 3.321 |
| 0.00 | 0.02 | 0.28 | 1.5 | 0 | 15 | 0.286 | 3.488 |
| 0.00 | 0.02 | 0.28 | 1.5 | 0 | 18 | 0.305 | 3.596 |
| 0.45 | 0.02 | 0.28 | 1.5 | B | 12 | 0.273 | 2.989 |
| 0.45 | 0.02 | 0.28 | 1.5 | B | 15 | 0.298 | 3.091 |
| 0.45 | 0.02 | 0.28 | 1.5 | B | 18 | 0.319 | 3.141 |
| 0.67 | 0.02 | 0.28 | 1.5 | C | 12 | 0.281 | 2.852 |
| 0.67 | 0.02 | 0.28 | 1.5 | C | 15 | 0.308 | 2.926 |
| 0.67 | 0.02 | 0.28 | 1.5 | C | 18 | 0.330 | 2.952 |
| 0.00 | 0.04 | 0.32 | 1.5 | 2 | 12 | 0.356 | 4.029 |
| 0.00 | 0.04 | 0.32 | 1.5 | 2 | 15 | 0.407 | 4.343 |
| 0.00 | 0.04 | 0.32 | 1.5 | 2 | 18 | 0.429 | 4.414 |
| 0.00 | 0.07 | 0.28 | 1.5 | 3 | 12 | 0.428 | 4.595 |
| 0.00 | 0.07 | 0.28 | 1.5 | 3 | 15 | 0.450 | 4.707 |
| 0.00 | 0.07 | 0.28 | 1.5 | 3 | 18 | 0.467 | 4.873 |
| 0.45 | 0.07 | 0.28 | 1.5 | 6 | 12 | 0.412 | 3.843 |
| 0.45 | 0.07 | 0.28 | 1.5 | 6 | 15 | 0.436 | 3.952 |
| 0.45 | 0.07 | 0.28 | 1.5 | 6 | 18 | 0.509 | 4.207 |
| 0.62 | 0.06 | 0.28 | 1.5 | 7 | 12 | 0.392 | 3.475 |
| 0.62 | 0.06 | 0.28 | 1.5 | 7 | 15 | 0.420 | 3.586 |
| 0.62 | 0.06 | 0.28 | 1.5 | 7 | 18 | 0.464 | 3.671 |
| 0.45 | 0.04 | 0.32 | 1.5 | 5 | 12 | 0.351 | 3.494 |
| 0.45 | 0.04 | 0.32 | 1.5 | 5 | 15 | 0.397 | 3.619 |
| 0.45 | 0.04 | 0.32 | 1.5 | 5 | 18 | 0.436 | 3.684 |
| 0.62 | 0.04 | 0.32 | 1.5 | A | 12 | 0.362 | 3.307 |
| 0.62 | 0.04 | 0.32 | 1.5 | A | 15 | 0.412 | 3.371 |
| 0.62 | 0.04 | 0.32 | 1.5 | A | 18 | 0.455 | 3.376 |
| 0.00 | 0.02 | 0.28 | 2.5 | - | 12 | 0.201 | 2.975 |
| 0.00 | 0.02 | 0.28 | 2.5 | - | 15 | 0.214 | 3.122 |
| 0.00 | 0.02 | 0.28 | 2.5 | - | 18 | 0.217 | 3.199 |
| 0.00 | 0.04 | 0.32 | 2.5 | - | 12 | 0.302 | 3.730 |
| 0.00 | 0.04 | 0.32 | 2.5 | - | 15 | 0.329 | 3.991 |
| 0.00 | 0.04 | 0.32 | 2.5 | - | 18 | 0.355 | 4.197 |
| 0.00 | 0.04 | 0.26 | 1.5 | 1 | 12 | 0.369 | 4.136 |
| 0.00 | 0.04 | 0.26 | 1.5 | 1 | 15 | 0.396 | 4.305 |
| 0.00 | 0.04 | 0.26 | 1.5 | 1 | 18 | 0.449 | 4.442 |
| 0.45 | 0.04 | 0.32 | 1.5 | 5H | 12 | 0.371 | 3.538 |
| 0.45 | 0.04 | 0.32 | 1.5 | 5H | 15 | 0.429 | 3.654 |
| 0.45 | 0.04 | 0.32 | 1.5 | 5H | 18 | 0.453 | 3.698 |

reason we have chosen values characteristic for a BHB, maximizing the effect on Mg$_2$ and
<Fe>. For Z=0.04 we have, apart from the model with Y=0.32, also calculated a model
with Y=0.26 (mixture 1), to investigate the effect of changing the He contents. The
integrated indices are hardly different between Y=0.26 and Y=0.32 (see also Table 4),
showing that Y is not a key parameter in determining Mg$_2$ and <Fe>. Next, we have
also investigated what happens if the mixing length to pressure scale height ratio $\alpha_{\mathrm{MLT}}$

is increased from 1.5 to 2.5. One sees that the models move to lower $Mg_2$ and $<Fe>$, again more or less parallel to the $Mg_2$ vs $<Fe>$ sequence. This means that in this way the metallicity that one can derive from integrated indices of galaxies can depend on the $\alpha_{MLT}$-parameter that one adopts. Finally, the change brought about by the use of the new OPAL opacities is very similar to the one by increasing $\alpha_{MLT}$. Interestingly, for the new opacities the calibration with a solar model leads to $\alpha_{MLT} \approx 1.6$. Therefore, the combination of opacities and calibrated $\alpha_{MLT}$ would result in similar line strengths for the example shown. Note that all dependencies we have discussed in this paragraph shift the line indices more or less parallel to the data along the lines of varying age. This implies firstly that the influence of especially age, $\alpha_{MLT}$, and total metallicity cannot be distangled by using $Mg_2$ and $<Fe>$ alone, and secondly that *only* the relative metal abundances give a substantial shift perpendicular to this direction into the region of observed values. For this reason the conclusion that [Mg/Fe] in elliptical galaxies is larger than solar is almost unavoidable.

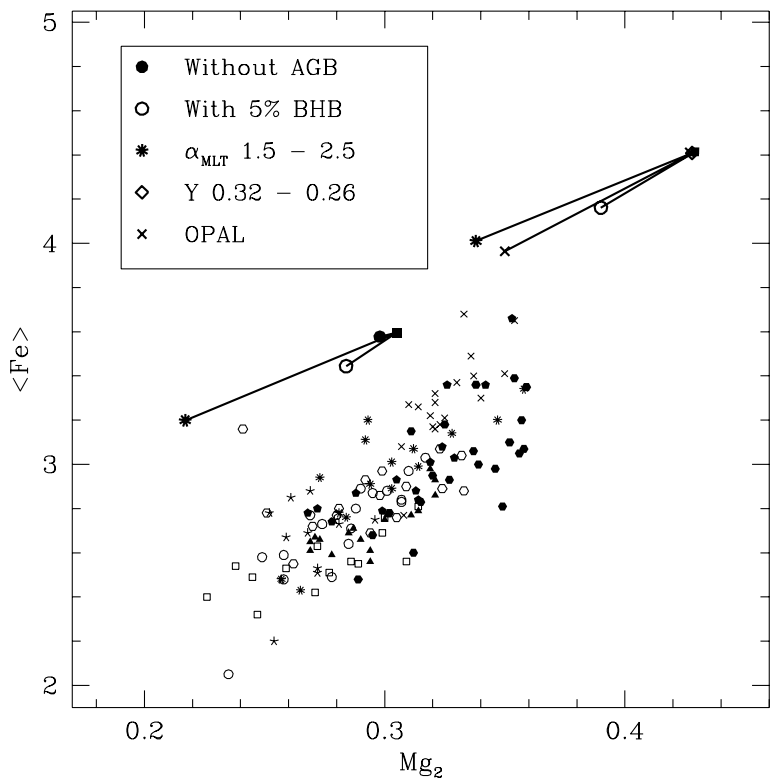

**Fig. 10.** Changes to the integrated indices as a result of changes of various ingredients of the stellar models (with solar metal ratios). For the description of the points see Fig. 8.

### 4.3 Comparison with observations of elliptical galaxies

Due to the fact that the new models use higher opacities they have redder turnoffs, implying that for a given metallicity and age the line indices $Mg_2$ and $<Fe>$ will be larger. If we use an age of 18 Gyr the range of the observed indices is fitted with $Z \lesssim 0.03$ for all galaxies (cf. Fig. 9), and the metallicities in the outer parts of some of the galaxies of Davies et al. (1993) are lower than $Z \approx 0.01$. If we assume an age of 12 Gyr, all galaxies are fitted with $Z \lesssim 0.04$. Since for the stellar models employed for Fig. 9 $\alpha_{MLT} = 1.5$ was used throughout, but the calibration with M67 (Fig. 6) and the Sun would yield a value close to 2, we would get higher metallicities for the ellipticals by using $\alpha_{MLT} = 2.0$. Use of the new OPAL opacities with the correspondingly lower $\alpha_{MLT}$ would not change this fit. It is clear that an average metallicity as large as $Z = 0.07$ is not required for any galaxy. With regard to [Mg/Fe] ratios: We find that the observed data yield [Mg/Fe] ratios between $\alpha_1$ (0.45) and $\alpha_2$ (0.67), both for low ($Z = 0.02$) as well as higher ($Z = 0.04$) total metallicities. Models with $\alpha_2$ seem to produce $<Fe>$ values that are slightly too low, although within the uncertainties they fit the data.

We can conclude here that we cannot fit the observations for $Mg_2$ and $<Fe>$ using state of the art stellar evolutionary models with solar element abundance ratios. Mg has to be enhanced with respect to Fe. Models with [Mg/Fe]=0.45 ($\alpha_1$) fit the data well, although the whole range in [Mg/Fe] between 0.3 and 0.7 cannot be excluded.

We obtain reasonable fits for giant elliptical galaxies with $Z = 0.02 - 0.04$. This implies that the iron metallicity (Renzini et al. 1993) is close to solar, and does not have to be super metal rich. It is worth comparing X-ray measurements of iron in the gas of some ellipticals with the inferred iron abundance in the stars. In particular, Serlemitsos et al. (1993) found an uncomfortably low iron abundance in NGC 1399 and NGC 4472 of respectively 0.56 and 0.20 $Fe_\odot$. However, NGC 1399 and NGC 4472 are the central galaxy in the Fornax cluster and the brightest galaxy in the Virgo cluster, resp. Therefore, dilution effects due to accretion of unenriched gas could play a rôle in determining the final abundances in the gas of these galaxies. On the other hand, Forman et al. (1994) recently found that in NGC 4472 the iron abundance is 1-2 times solar. Although this kind of measurements look still quite uncertain, this result is in better agreement with the predictions of our model, where the iron abundance in the gas restored from dying stars in ellipticals is expected to be higher than in the stars and the [Mg/Fe] ratio is expected to be negative (see also Renzini et al. 1993). An iron metallicity similar to or higher than that of the stars is also found in quiescent spiral galaxies (Balcells & Peletier 1994).

### 4.4 Implication for formation models of elliptical galaxies

Any conclusion on the [Mg/Fe] ratio in stars of elliptical galaxies is relevant to the understanding of the mechanism of formation of such objects. In fact, abundance ratios represent a powerful constraint on the galaxy formation process, since they depend mostly on the assumed nucleosynthesis prescriptions and stellar progenitors. In particular, the stellar progenitors constrain the timescale for the first appearance of a chemical element into the ISM. As mentioned in the introduction, abundance ratios have already been used to impose constraints on the formation mechanism of our own Galaxy. From the observed behaviour of the [$\alpha$/Fe] ratios in the solar vicinity one can infer the timescale for the process of formation of the Galactic halo (see Matteucci & François 1992; Smecker-Hane & Wyse 1992). To calculate this, one needs to assume that SNe of type II are responsible for the bulk of $\alpha$-elements whereas type Ia SNe are responsible for the bulk of Fe, and also that the initial mass function (IMF) does not vary with time. In this framework, the fact that the predominant stellar population in elliptical galaxies has a solar or supersolar metallicity and overabundances of $\alpha$-elements to Fe relative to the solar value indicates that most of the iron we observe in ellipticals has been formed in type II SNe. As a consequence of this, the process of formation of these objects must have been much faster (e.g. more efficient star formation) than in the solar neighbourhood, as was already predicted by Matteucci & Brocato (1990).

Models of chemical evolution of elliptical galaxies have recently been discussed by Matteucci (1992b, 1993). She discussed the implications of the observed <Fe> vs. Mg$_2$ relation for nuclei of ellipticals, in particular the evidence for overabundances of $\alpha$-elements and the fact that the [$\alpha$/Fe] ratio seems to increase with galactic mass. This last implication, which arises from the flat slope of the Mg$_2$ – <Fe> relation for nuclei of galaxies – as opposed to the relation within a galaxy– is in contradiction with predictions of Matteucci & Tornambè (1987), who suggested that Fe should have increased more than Mg when increasing the galactic mass. The model of Matteucci & Tornambè (1987) is very similar to that of Arimoto & Yoshii (1987), the main difference being that in Matteucci & Tornambè the evolution of Fe is followed in detail through the introduction of SNe of type Ia. One basic assumption of the model is that the efficiency of star formation is a decreasing function of the galactic mass (see discussion in Matteucci 1993).

Under these assumptions Matteucci & Tornambè (1987) predicted that the time of occurance of a galactic wind is an increasing function of the galactic mass, and that [Mg/Fe] is a decreasing function of galactic mass. So, to understand the trend of [Mg/Fe] observed in bright ellipticals one should change some of the assumptions made in this model and invoke either an increasing star formation efficiency with galactic mass, creating the situation where galactic winds occur sooner in massive than in smaller ellipticals, or a variable IMF from galaxy to galaxy, to explain the observed trend (Matteucci 1993). The variation of the IMF should be substantial and go in the sense of favoring more massive stars in more massive galaxies. One could also achieve the situation of earlier winds in more massive objects if the amount and distribution of dark matter varies from galaxy to galaxy. This possibility however has not been quantitatively explored.

From a direct comparison of Matteucci's (1993) model with the results of the previous section we can impose constraints on the timescale for the formation of ellipticals. In Fig. 11 we show the predicted [Mg/Fe] ratio in the gas of an elliptical galaxy with initial luminous mass of $10^{11} M_\odot$ and a Salpeter (1955) IMF. This figure shows that [Mg/Fe] $\geq 0$ when t $\leq 3\,10^8$ yrs and this result does not change by assuming a flatter IMF than the Salpeter one.

To conclude, the results from the previous section (i.e. that the average [Mg/Fe] in ellipticals should lie between +0.3 and +0.7 dex), show that the formation of the bulk of stars in elliptical galaxies cannot have lasted more than $\simeq 3 \cdot 10^8$ years, for any reasonable IMF which is assumed to be constant in time. This timescale corresponds to the time after which the contribution to the chemical enrichment from type Ia SNe is not negligible any more. This timescale is universal since it depends only on the lifetimes of the assumed progenitors for type Ia SNe, namely white dwarfs in binary systems (see Matteucci & François 1992).

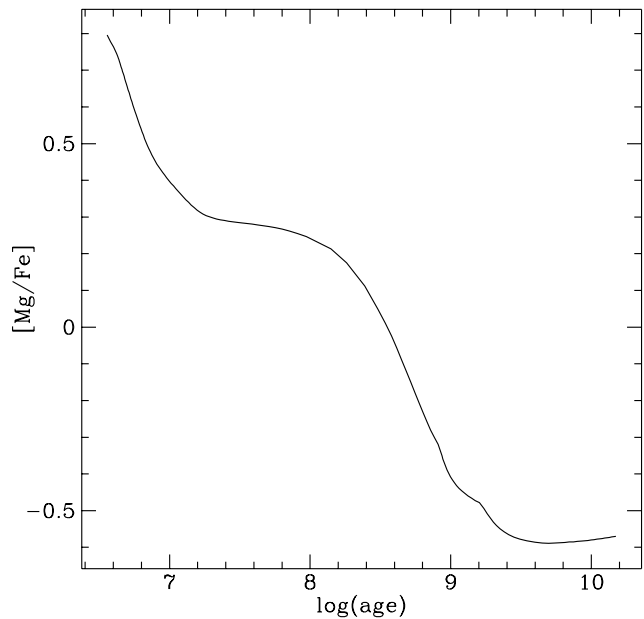

**Fig. 11.** Predicted [Mg/Fe] ratio in the gas of a galaxy with initial luminous mass $10^{11} M_\odot$ and a Salpeter IMF as a function of the logarithm of the galactic age (expressed in years).

# 5 Conclusions

The main conclusions of this paper are as follows:

We presented new stellar evolutionary models for old stars of metallicities larger than solar and with non-solar abundance ratios. The models have been calculated using the program of Weiss (1989) with several modifications, including a better treatment of the evolution on the RGB, and the explicit inclusion of H, He, C, N and O in the equation of state, the opacities and in a nuclear reaction network. Stellar models were calculated in the mass range of $M = 0.6 \ldots 1.1 M_\odot$ and 8 chemical mixtures with $Z = 0.02, 0.04, 0.06,$ and 0.07, where the metals had either solar or $\alpha$-enhanced relative abundances.

We have applied the new models to calculate synthetic indices of lines of Mg and Fe in elliptical galaxies to quantify the results by Peletier (1989) and Worthey et al. (1992) that Mg is overabundant with respect to Fe in bright elliptical galaxies. Our models are able to explain the observations if [Mg/Fe] is between 0.3 and 0.7 (best value $\sim 0.45$). Solar abundance ratios are excluded. We find that changes in the age of the system, the assumed mixing length parameter, the total metallicity, or the inclusion of an AGB/HB population all lead to changes in the $Mg_2$ and $<Fe>$ line strengths that are parallel in the $Mg_2 - <Fe>$ plane, and as such do not affect our result that [Mg/Fe] in ellipticals is larger than solar. Neither can they be determined independently. We obtain the best fit for the centers of elliptical galaxies with $Z = 0.04$ (total metal content), $\alpha_{MLT} = 2.0$ and an old age (15-18 Gyr), although these values may depend on the input models. Due to the enhancement in $\alpha$-elements the iron abundance is practically solar.

We have also demonstrated that it is possible to use stellar models with solar metal ratios for the calculation of $\alpha$-enhanced metal line strengths, if one assumes $\alpha$-enhanced metals for the line indices (cf. models 5 and 5H in Fig. 10) and the total metallicity Z is the same. Although the results using this simplified approach do not exactly match those obtained with the fully consistent stellar models, the approximation is a very good one and sufficient at the present stage.

For the formation of elliptical galaxies our previous result shows that the bulk of the stars in a bright elliptical galaxy must have formed within a time scale of $3 \times 10^8$ years, for any reasonable IMF assumed to be constant in time (Matteucci 1993). This time scale is given by the relative frequency of SNe Ia and SNe II. A high value of [Mg/Fe] implies that most of the heavy element enrichment must have occurred through SNe II, and that the bulk of the stars must have formed on short timescales compared to that of SNe Ia's. The high [Mg/Fe] implies that the bulk of the material in the inner parts consists of stars that are more or less coeval, and these results speak against e.g. merging scenarios of two spiral galaxies in which a significant part of the stars in the center of the resulting elliptical is formed during the merger.

*Acknowledgements*. We thank E. Brocato for helpful discussions and advice in the early stages of this project. We acknowledge the efforts of E. Valentijn and T.S. van Albada during the course of R. Peletier's Ph.D. project, when part of this work was done. One of us (F.M.) wants to thank S. Pellegrini for pointing out the new results by Forman and coworkers. A.W. thanks A. Tornambè for pointing out to him the importance of $\alpha$-elements for stellar models.